\documentclass[aps,prb,twocolumn,showpacs]{revtex4}
\usepackage{latexsym}
\usepackage{amssymb}
\usepackage{graphicx}
\begin{document}
\title{Disorder effects on the spin-Hall current
in a diffusive Rashba two-dimensional heavy-hole system}
\author{S. Y. Liu}
\email{liusy@mail.sjtu.edu.cn}
\author{X. L. Lei}
\affiliation{Department of Physics, Shanghai Jiaotong University, 1954
Huashan Road, Shanghai 200030, China}
\date{\today}
\begin{abstract}
We investigate the spin-Hall effect in a two-dimensional
heavy-hole system with Rashba spin-orbit coupling using a
nonequilibrium Green's function approach. Both the short-
and long-range disorder scatterings are considered in
the self-consistent Born approximation. We find that, in the case of long-range collisions, the
disorder-mediated process leads to an enhancement of the spin-Hall
current at high heavy-hole density, whereas for short-range scatterings
it gives a vanishing contribution.
This result suggests that the recently observed spin-Hall
effect in experiment is a result of the
sum of the intrinsic and disorder-mediated contributions. We have
also calculated the temperature dependence of spin-Hall
conductivity, which reveals a decrease with increasing the temperature.
\end{abstract}

\maketitle

\section{Introduction}

Spin-Hall effect has been extensively studied in semiconductors
with spin-orbit (SO) coupling due to its potential applications in the
emerging field of spintronics. It
refers to an appearance of the net spin current flowing perpendicular
to the applied dc field. This transport phenomenon was
studied by Dyakonov
and Perel in 1971\cite{DP} and by Hirsch more
recently.\cite{HS} Therein, the authors demonstrated that the
spin-dependent scattering between electrons and impurities causes
the electrons with opposite spins to veer to different sides of
the considered systems. Obviously, such spin-Hall effect strongly
relies on the electron-disorder collision and hence, lately, has
been named as extrinsic spin-Hall effect.\cite{Sinova}

Recently, an impurity-free spin-Hall effect, namely
intrinsic spin-Hall effect, has been proposed by Sinova {\it et
al.} in two-dimensional (2D) electron systems\cite{Sinova} and
Murakami {\it et al.} in p-doped bulk
semiconductors,\cite{Murakami} respectively. This spin-Hall effect
is associated with the dc-field-induced transitions between the
spin-orbit-coupled bands and contributes from
all occupied electron states below the Fermi energy.\cite{Liu1} In
2D electron semiconductors with Rashba and/or Dresselhaus
spin-orbit coupling, the value of intrinsic spin-Hall conductivity
has been found to be a universal value $e/8\pi$.\cite{Sinova,Shen}
Further, in 2D Rashba heavy-hole (HH) systems, Schliemann and Loss
reported that the intrinsic spin-Hall conductivity starts out at
value $9e/8\pi$ and increases with increasing the Rashba
coupling.\cite{Schliemann}

However, the disorder can strongly affect the spin-Hall effect,
especially in 2D semiconductors.\cite{Liu1,Schliemann1,
Burkov,Nomura, Inoue, Mishchenko, Dimitrova, Chalaev,
Khaetskii,Chao,Liu2,Bauer} There exists another collision-related
spin-Hall current connecting to the electron states near the Fermi
surface.\cite{Liu1} It is disorder-mediated but independent of the
impurity density. In diffusive Rashba 2D electron systems, this
disorder-mediated spin-Hall effect leads to a complete cancellation
of the total spin-Hall current for both the short- (Refs. \onlinecite{Inoue,
Mishchenko, Dimitrova, Chalaev, Khaetskii}) and long-range
disorders.\cite{Liu1} The same conclusion has been obtained even in
the presence of both the linear in momentum Rashba and Dresselhaus
SO couplings.\cite{Liu2,Chao,Bauer}

Experimentally, observations of spin-Hall effect have been
reported recently by Kato {\it et al.}\cite{Kato} and by Wunderlich
{\it et al.}\cite{Wunderlich} The latter work is of special
interest, since there the spin-Hall effect has been found in a
relatively {\it clean} 2D heavy-hole sample with Rashba spin-orbit
coupling. Obviously, in such a system the observed spin-Hall effect cannot
be understood as an extrinsic one. To interpret this
experimental finding, it immediately arises the question whether
the sum of the intrinsic and the
disorder-mediated contributions to spin-Hall current also vanishes in
the Rashba 2D heavy-hole semiconductors. By means of Kubo formalism, Bernevig and Zhang
found that the disorder has no effect on the spin-Hall effect
and the observed spin-Hall effect is a pure
intrinsic one.\cite{Bernevig}

Based on a nonequilibrium kinetic equation approach, we demonstrate in this paper
that the vanishing of the disorder effect on spin-Hall current in 2D HH systems occurs only
for short-range hole-impurity scattering. When the disorder becomes long-ranged,
the contribution from disorder-mediated process to the
spin-Hall current can not be ignored.
It has the same sign as the intrinsic one at
high HH density, but opposite for low density.
The comparison between theory and experiment indicates that
the observed spin-Hall effect is the result of the sum of the
disorder-mediated and intrinsic contributions.
We also find that the spin-Hall conductivity decreases
with the rise of the temperature.

The paper is organized as follows. In Sec. II
the noninteracting Green's functions in spin basis and the kinetic
equations for nonequilibrium distribution functions are presented. In Sec. III
we analyse the disorder effects on spin-Hall conductivity for short- and
long-range hole-impurity scatterings, respectively.
Finally, we conclude our results in Sec. IV.

\section{Formalism}

In bulk semiconductors like GaAs, the heavy- and light-hole (LH)
bands with total angular momentum $j=3/2$ are degenerate at the
band edge. However, in quasi-two-dimensional semiconductors,
the additional confinement along the growth direction
of heterostructure yields
a splitting between these bands and the degeneracy is lifted. As a
result, the HH and LH subbands are separated.

We consider a two-dimensional HH system, where only the lowest
heavy-hole subband is occupied. Note that this condition can be satisfied
when the quasi-2D system is sufficiently narrow and the density and the temperature
are not too high. In this way, the effective Hamiltonian
for a single heavy-hole subjected a spin-orbit
interaction due to the structural inversion asymmetry, can be written
as\cite{Schliemann,Bir,GS,Winkler,Winkler1,Winkler2, Add}
\begin{equation}
{\hat{\bar H}}=\frac{{\bf p}^2}{2m}+i\frac{\alpha}{2}(p_-^3{\hat \sigma}_+-p_+^3
{\hat \sigma}_-),\label{Ham}
\end{equation}
with 2D momentum ${\bf p}\equiv (p\cos
\phi_{\bf p},p\sin \phi_{\bf p})$,
${\hat \sigma}_\pm\equiv {\hat \sigma}_x\pm {\hat \sigma}_y$, and $p_\pm\equiv
p_x\pm i p_y $. ${\hat {\bf \sigma}}\equiv ({\hat \sigma}_x,{\hat \sigma}_y,{\hat \sigma}_z)$
are the Pauli matrices and $m$ is the
hole effective mass. Note that this Hamiltonian is obtained
through the decomposition of the HH-LH coupling by third-order
L\"owdin perturbation theory.\cite{Bir} The parameter $\alpha$ is
proportional to the Rashba spin-orbit coupling constant, and
relies on the separation between
HH and LH subbands and the Luttinger parameters.\cite{Winkler,Winkler1} We
can see from Eq.\,(\ref{Ham}) that the spin-orbit interaction is cubic
in momentum. There exists another type of the cubic SO coupling, namely
the Dresselhaus spin-orbit coupling in 2D electron systems. The
spin-Hall effect for such SO interaction has already been
studied in Ref.\,\onlinecite{Chao}.

It follows from Eq.\,(\ref{Ham}) that
the Green's function of noninteracting particles has the form
\begin{equation}
\hat {\bar {\rm G}}_0^{r,a}(\omega,p)={\hat \Pi}^{(1)}{\rm G}_{01}^{r,a}(\omega,p)+
{\hat \Pi}^{(2)}{\rm G}_{02}^{r,a}(\omega,p),
\end{equation}
with ${\rm G}_{0\mu}^{r,a}=[\omega-\varepsilon_\mu(p)\pm i\delta]^{-1}$
($\mu=1,2$).
Here,
\begin{equation}
\varepsilon_{\mu}( p)=\frac{ p^2}{2m}+(-1)^\mu \varepsilon_{HH},
\end{equation}
\begin{equation}
\hat {\Pi}^{(\mu)}_{\alpha\beta}=\frac 12\left [\delta_{\alpha\beta}+\frac{(-1)^{\mu}}{p}
({ p}_-^3({\hat \sigma}_+)_{\alpha\beta}-{p}_+^3
({\hat \sigma}_-)_{\alpha\beta})\right ],
\end{equation}
and $\varepsilon_{HH}\equiv \alpha p^3$.
The operator ${\hat \Pi}^{\mu}$ represents the projection onto the states
with a definite helicity.

By means of a local unitary transformation
\begin{equation}
{\hat U}(\bf p)=\frac 1 {\sqrt{2}}\left (
\begin{array}{cc}
1&1\\
i{\rm e}^{3i\phi_{\bf p}}&-i{\rm e}^{3i\phi_{\bf p}}
\end{array}
\right ),\label{Uni}
\end{equation}
the Hamiltonian (\ref{Ham}) can be diagonalized as ${\hat H}\equiv {\hat U}^+{\hat {\bar H}}{\hat U}={\rm
diag}(\varepsilon_{1}(p), \varepsilon_{2}( p))$, and hence
the noninteracting Green's function also becomes diagonal $\hat {\rm
G}_0^{r,a}\equiv {\hat U}^+{\hat {\bar {\rm
G}}_0^{r,a}}{\hat U}={\rm diag}({\rm G}_{01}^{r,a},{\rm G}_{02}^{r,a})$.
This implies that the spin-orbit interaction
can lead to further lifting the degeneracy of the HH band. In result,
two spin-orbit-coupled bands with dispersion relations
$\varepsilon_{\mu}(p)$ are formed.
Note that above transformation corresponds to a change of the basis of
the Hilbert space from a spin to a helicity one.
In the following study, the spin-Hall effect will be carried out
in the helicity basis.

We assume a weak dc field ${\bf E}$ applied to the 2D HH systems along the $x$
direction. We are interested in the
$z$-direction-polarized spin current flowing along the $y$
direction, {\it i. e.} the spin-Hall current. It is known that the
single-particle spin current operator can be defined
by\cite{Schliemann}
\begin{equation}
{j}_y^z=\frac{3}{2}\frac 1{2e}(
j_y{\hat \sigma}_z+{\hat\sigma}_zj_y),
\end{equation}
where the factor $\frac{3}{2}$ reflects the angular momentum of
the heavy-hole and the $j_y$ is the electric current operator.
By taking the unitary transformation,
$j^z_y$ becomes nondiagonal and the net observed
spin-Hall current can be expressed as
\begin{equation}
{J}^{z}_y=\sum_{{\bf p}}\frac {3p_y}{2m}[{\hat \rho}_{12}\left
({\bf p})+{\hat \rho}_{21}({\bf p})\right ].\label{Jz}
\end{equation}
The ${\hat \rho}_{\alpha\beta}$ ($\alpha,\beta=1,2$) are the elements of
the distribution function ${\hat \rho}({\bf p
})$: ${\hat \rho}_{\alpha \beta}({\bf p
})=-i\int \frac{{\rm d}\omega}{2\pi} \hat {\rm G}_{\alpha \beta}^<({\bf
p},\omega)$ with $\hat {\rm G}^<$ being the less Green's function.
The spin-Hall conductivity is given by $\sigma_{sH}=J^y_z/E$.

To carry out the spin-Hall conductivity, it is necessary to derive the
kinetic equation for less Green's function. Following the
procedures described in Ref.\,\onlinecite{Liu1}, we find
that $\hat \rho$ obeys the equation
\begin{widetext}
\begin{equation}
e{\bf E}\cdot \nabla_{\bf
p}{\hat \rho}({\bf p}) +\frac {3ie {\bf E}}{2}\cdot \nabla_{\bf
p} \phi_{\bf p}
[{\hat \rho},{\hat \sigma}_x]+i\varepsilon_{HH}[{\hat \rho},{\hat \sigma}_z]=-\left .\frac
{\partial {\hat \rho}}{\partial T}\right |_{\rm scatt}.\label{KE}
\end{equation}
At the same time, the self-energies for the hole-impurity interaction with
isotropic scattering matrix $V(|{\bf p}-{\bf k}|)$ can be written as
\begin{equation}
{\hat \Sigma}^{r,<}({\bf p},\omega)=\frac 12 n_i\sum_{{\bf k}}|V(|{\bf
p}-{\bf k}|)|^2\left \{ a_1 \hat {\rm G}^{r,<}+a_2\hat\sigma_x\hat{\rm
G}^{r,<}\hat\sigma_x+ia_3[\hat\sigma_x,\hat{\rm G}^{r,<}]\right \},
\end{equation}
\end{widetext}
with $n_i$ being the impurity density and $a_i$($i=1,2,3$) the factors associated with the
directions of the momenta, $a_1=1+\cos (3\phi_{\bf p}-3\phi_{\bf
k})$, $a_2=1-\cos (3\phi_{\bf p}-3\phi_{\bf k})$, $a_3=\sin
(3\phi_{\bf p}-3\phi_{\bf k})$. The relaxation term $\left
.\frac{\partial \hat\rho}{\partial T}\right |_{\rm scatt}$ takes a
general form.\cite{Liu2}

Similar to the case of 2D electron systems, the linearized kinetic
equation can be further simplified and its solution is found to
comprise two terms. The first one is
\begin{equation}
\hat\rho_{12}^{(1)}({\bf p})=\hat\rho_{21}^{(1)}({\bf p})=-\frac{3e{\bf
E}\cdot {\nabla_{\bf p}\phi_{\bf p}}}{4\varepsilon_{HH}}\left
\{n_{\rm F}[\varepsilon_1( p)]-n_{\rm F}[\varepsilon_2( p)]\right
\},
\end{equation}
and the real part of the second one takes the form
${\rm Re} \hat\rho^{(2)}_{12}({\bf p})=\zeta (p)eE\sin \phi_{\bf p}$
with function $\zeta(p)$
\begin{widetext}
\begin{eqnarray}
\zeta (p) &=&\frac 1{4\alpha p^3} \sum_{\mu=1,2}\left \{ \frac
{(-1)^\mu}{\tau_{4\mu\mu}}\left . \frac{\partial n_{\rm
F}(E)}{\partial E}\right |_{E=\varepsilon_\mu(p)}
\Phi_\mu[\varepsilon_\mu(p)]- \frac
{(-1)^\mu}{\tau_{4\mu\bar{\mu}}}\left . \frac{\partial n_{\rm
F}(E)}{\partial E}\right |_{E=\varepsilon_{{\mu}}(p)} \Phi_{\bar
\mu}[\varepsilon_\mu(p)]\right \}.\label{ZZZ}
\end{eqnarray}
\end{widetext}
The functions $\Phi_\mu(E)$
connect to the diagonal distribution functions
\begin{equation}
\rho^{(2)} _{\mu\mu} ({\bf p})=-\left .\frac{\partial n_{\rm F}
(E)}{\partial E} \right
|_{E=\varepsilon_\mu(p)}\Phi_{\mu}[\varepsilon_\mu(p)]eE\cos
\phi_{\bf p}
\end{equation}
and can be given by the coupled
equations,
\begin{equation}
\frac {\partial \varepsilon_\mu (p)}{\partial p}=
\frac{\Phi_\mu[\varepsilon_\mu(p)]}{\tau_{1\mu\mu}}
+\frac{\Phi_\mu [\varepsilon_\mu(p)]}{\tau_{2\mu\bar{\mu}}}
-\frac{\Phi_{\bar{\mu}}
[\varepsilon_\mu(p)]}{\tau_{3\mu\bar{\mu}}},\label{EQR1}
\end{equation}
with $\bar{\mu}=3-\mu$. In these equations, $\tau_{i\mu\nu}$
denote the different relaxation times
\begin{equation}
\frac {1}{\tau_{i\mu\nu}}=2\pi n_i\sum_k |V(|{\bf p}-{\bf k}|)|^2
\Lambda_{i\mu\nu} (\phi_{\bf k} -\phi_{\bf p},p,k),
\end{equation}
where the functions $\Lambda_{i\mu\nu}(\phi,p,k)$ are
defined as $\Lambda_{1\mu\nu}(\phi,p,k)=\frac 12
[1+\cos(3\phi)](1-\cos \phi) \delta (\varepsilon _{\mu
p}-\varepsilon _{\nu k})$, $\Lambda_{2\mu\nu}(\phi,p,k)=\frac 12
[1-\cos(3 \phi)] \delta (\varepsilon _{\mu p}-\varepsilon _{\nu
k})$, $\Lambda_{3\mu\nu}(\phi,p,k)=\frac 12 \cos \phi[1-\cos(3
\phi)] \delta (\varepsilon _{\mu p}-\varepsilon _{\nu k})$ and
$\Lambda_{4\mu\nu}(\phi,p,k)=\frac 12 \sin \phi\sin(3 \phi) \delta
(\varepsilon _{\mu p}-\varepsilon _{\nu k})$.

The first term in the solution of the linearized kinetic equation gives rise
to an intrinsic spin-Hall conductivity $\sigma_{sH}^{(1)}$
\begin{equation}
\sigma_{sH}^{(1)}=\frac{-9e}{16\pi
m\alpha}\int_0^{\infty}\frac{{\rm d}p}{p^2} \{n_{\rm
F}[\varepsilon_1(p)-\mu]-n_{\rm F}[\varepsilon_2(p)-\mu]\},\label{HHH}
\end{equation}
in agreement with the previous
study.\cite{Schliemann} This part of spin-Hall
conductivity comes from the interband transition processes
in which all holes below Fermi surface join. It is associated with
the energy separation between two spin-orbit-coupled bands in
helicity basis and reveals an intrinsic character:
it is independent of any hole-impurity scattering. Physically, the dc field
can cause an elastic transition of a hole from one band to another
one when this hole gains an additional momentum from the external
dc field.

The second term in the solution is associated with the transport
process and only the hole states near the Fermi
surface contribute. It relates to the disorder collision in a surprising way:
its contribution to the spin-Hall conductivity $\sigma_{sH}^{(2)}$ is independent of the
impurity density $n_i$
but depends on the above-defined relaxation times and hence the form of
scattering matrix. We can understand the origin of the $\sigma_{sH}^{(2)}$ as follows.
When an external dc field is applied,
the holes should participate in the longitudinal transport,
leading to a diagonal distribution proportional to the inverse of the impurity density.
At the same time, these perturbative holes also
experience the impurity scattering, yielding the interband polarization.
In result, the nondiagonal distribution
becomes impurity-density-independent. Here the impurity plays only an
intermediate role.

We should note that this disorder-mediated mechanism is physically identical with
the side-jump mechanism studied recently in the investigation on
spin-Hall effect.\cite{SJ} Contributions
to spin-Hall conductivity from both mechanisms are independent
of the impurity density but collision-related. At the same time,
they all are associated with the hole (or electron) states near the
Fermi surface. However, formally, these two mechanisms look
different because of two different approaches. The side-jump process is carried out for a
spin-orbit interaction involved in the electron-impurity scattering.
It corresponds to a lateral displacement of the center of the
wave-packet during the scattering and hence connects to the
scattering-dependent term of the current operator. However, in our
study, since the spin-orbit coupling is included in the free-hole (or
free-electron) Hamiltonian, the current operator is independent
of the hole-impurity collision.

The fact that the spin-Hall current consists of two parts, is
similar to the well-known result of St\v reda,\cite{Streda}
as well as the recently obtained conclusion in the context of
anomalous Hall effect (AHE).\cite{AHE} In the 2D electron systems
magnetically or with magnetization, the off-diagonal conductivity
usually comes from two terms, one of which is due to the electron
states near the Fermi energy and the another one is related to the
contribution of all occupied electron states below the Fermi
energy.

\section{Results and discussions}

First, we consider the spin-Hall effect for the short-range disorder.
In this case, the hole-impurity collision matrix
has a simple momentum-independent form $V(|{\bf p}-{\bf k}|)\equiv
u$ and the scattering is described by a single relaxation
time $\tau=1/n_i u^2$. It can be seen
from Eq.\,(\ref{ZZZ}) that the disorder-mediated contribution to spin-Hall current
relies on function $\Lambda_{4\mu\nu}$ proportional to $\sin\phi\sin(3\phi)$.
When the angle integration in Eq.\,(\ref{Jz}) is performed, the vanishing $\sigma_{sH}^{(2)}$
is obtained.

This result agrees with that of Ref.\,\onlinecite{Bernevig},
in which the vanishing of disorder-related spin-Hall conductivity
comes from an analogous angle-integration. However, for 2D electrons with Rashba spin-orbit coupling,
the function $\Lambda_{4\mu\nu}$ depends on $\sin^2\phi$, leading to a nonzero $\sigma_{sH}^{(2)}$.\cite{Liu1}

For long-range disorders, the additional momentum-dependence of
scattering matrix produces rich novel phenomena. We have performed
a numerical calculation to investigate the long-range disorder
effect on the spin-Hall conductivity in a 2D GaAs/AlGaAs based
heavy-hole system. Consider a Coulomb interaction
between the 2D heavy-holes and the charged impurities located
at a distance $s=500$\,\AA\,\, from the 2D plane:
$V(p)\sim{\rm e}^{-sp}I(p)$.\cite{Ando}
$I(p)$ is the form factor. In calculation, we take
the effective mass $m=0.27m_{\rm e}$ (Ref. \onlinecite{Wunderlich}) and
the coupling constant $m\alpha=5$\,\AA.

In Fig.\,1 the total and disorder-mediated spin-Hall
conductivities are plotted as functions of heavy-hole density at zero
temperature. The intrinsic spin-Hall conductivity is
almost independent of HH density and has the value $9e/8\pi$,
in agreement with the result in Ref.\,\onlinecite{Schliemann}. It is
evident from Fig.\,1 that, unlike the case of short-range
disorder, here the $\sigma_{sH}^{(2)}$ and $\sigma_{sH}$ decrease
with descending HH density. $\sigma_{sH}^{(2)}$ even
becomes negative for $n_p<2.5\times 10^{10}$\,cm$^{-2}$ due to the sign change
of the quantities $\tau_{4\mu\nu}$.

\begin{figure}
\includegraphics [width=0.45\textwidth,clip] {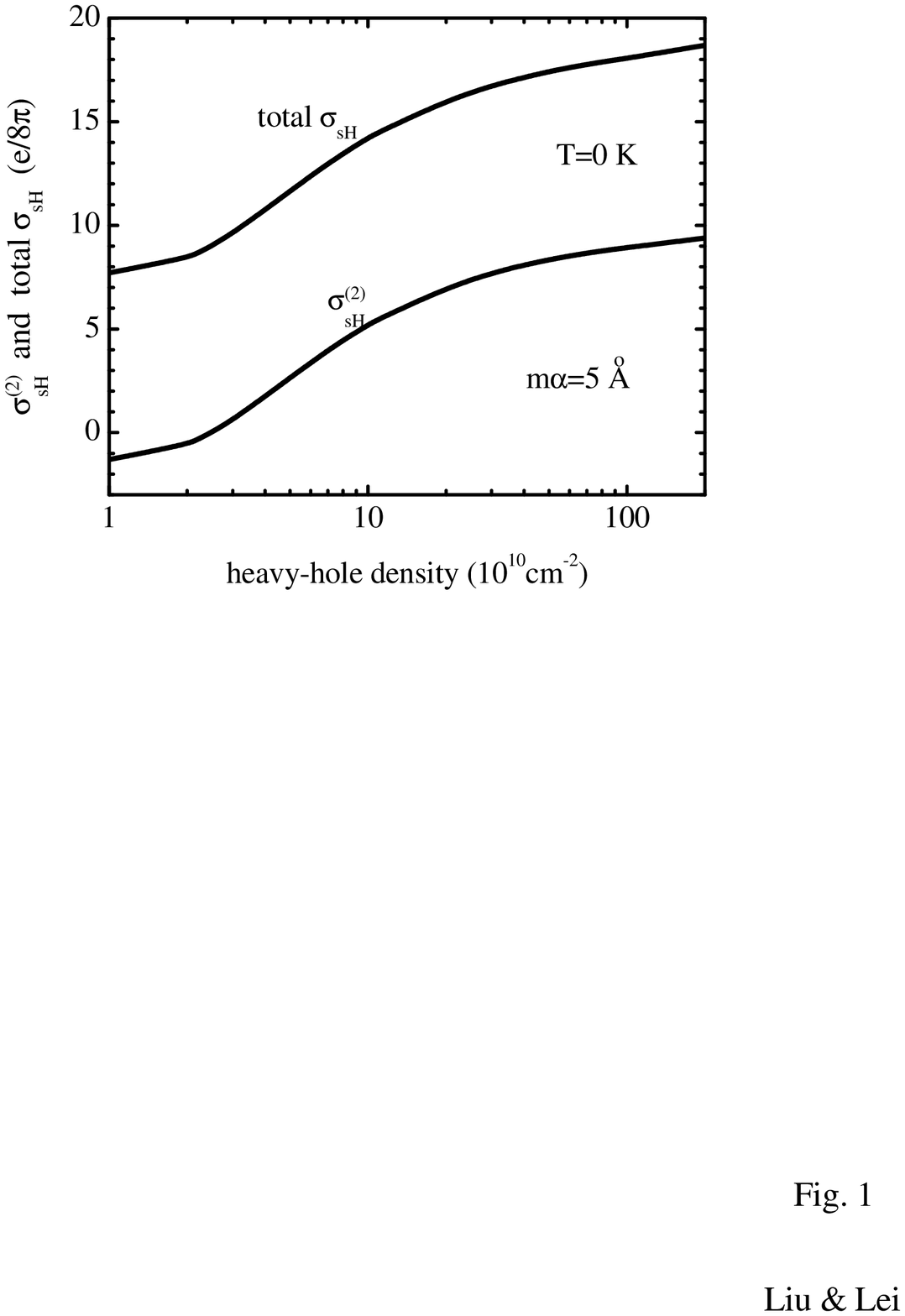}
\caption{Density dependence of disorder-mediated and total
spin-Hall conductivity in a Rashba 2D heavy-hole system. The
lattice temperature $T=0$\,K and the coupling constant
$m\alpha=5$\,\AA. } \label{fig1}
\end{figure}

For HH density of order of $10^{10}$\,cm$^{-2}$, the Fermi energy
and the temperature (less than $4.2$\,K) become comparable. Hence,
the HH spin-Hall effect should be strongly influenced by temperature.
The calculation indicates that with ascending the temperature, the
intrinsic spin-Hall conductivity begins with the zero-temperature value $9e/8\pi$
and descends for heavy-hole densities
$n_p$ in the range $1-100\times 10^{10}$\,cm$^{-2}$. However, as
shown in Fig.\,2(a), for the disorder-mediated spin-Hall
conductivity, there exists a critical HH density $n_{pc}$
about equal to $4\times 10^{10}$\,cm$^{-2}$. With a rise
of the temperature, the $\sigma_{sH}^{(2)}$ increases for
$n_p<n_{pc}$, while it decreases for the opposite case. However,
one can see from Fig.\,2(b) that the total $\sigma_{sH}$ always
decreases with increasing temperature.
This implies that the descending intrinsic spin-Hall
conductivity plays a dominant role at low HH density.

\begin{figure}
\includegraphics [width=0.45\textwidth,clip] {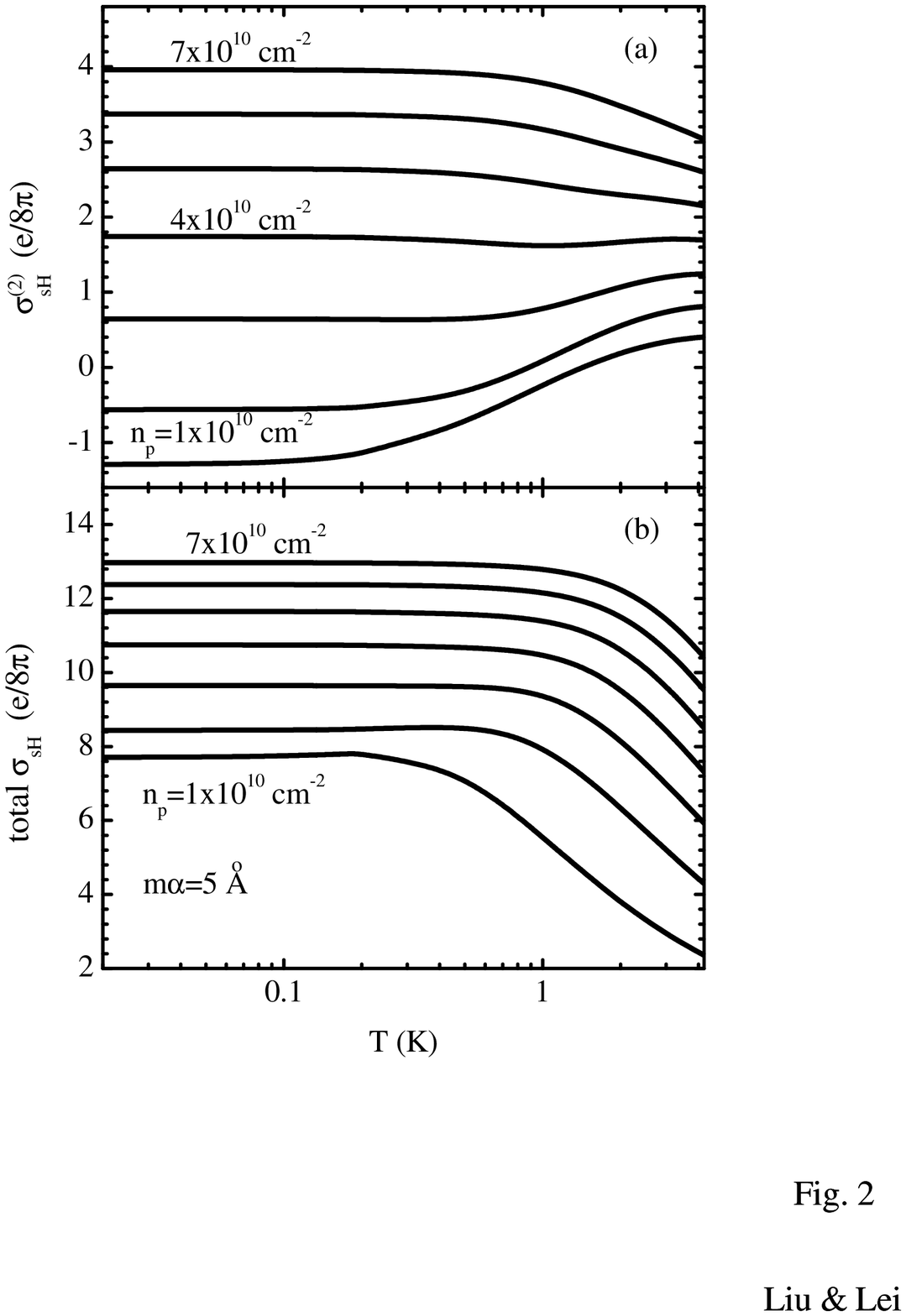}
\caption{Temperature dependence of disorder-mediated and total
spin-Hall conductivity in a Rashba 2D heavy-hole system. From
bottom to top, the corresponding HH densities are $n_p=1$, $2$,
$3$, $4$, $5$, $6$, and $7\times 10^{10}$\,cm$^{-2}$,
respectively. The coupling constant $m\alpha=5$\,\AA.}
\label{fig2}
\end{figure}

Experimental study by Wunderlich {\it et al.} has been performed
for heavy-hole density $n_{p}=2\times 10^{12}$\,cm$^{-2}$.
\cite{Wunderlich} At such high HH density, the value of spin-Hall
conductivity remains almost unchanged for temperatures less than
$4.2$\,K. At the same time, the disorder-mediated process plays a
positive role and its contribution to spin-Hall conductivity adds
to the intrinsic one. Hence, the 2D HH systems with such a high HH
density serve as good candidates for the observation of the spin-Hall
effect.

To compare with the experiment, we have calculated the spin-Hall
conductivity for the HH system with hole density equal to the
experimental one. The Fermi energy is approximately
chosen to be $17.5$\,meV and hence the coupling constant is
$m\alpha=2.5$\AA. At $T=4.2$\,K, we obtain the total spin-Hall
conductivity $\sigma_{sH}=18.2e/8\pi$. This result is independent
of the impurity density, but relies on the forms of the scattering
matrix. Although the spin-Hall conductivity is not a
quantity directly observed in experiments, our investigation suggests that
the observed spin-Hall effect may not only be a pure intrinsic one, as
claimed in Ref.\,\onlinecite{Bernevig}, rather comes from both the
intrinsic and the disorder-mediated processes.

\section{Conclusion}

The spin-Hall effect of a 2D heavy-hole system with Rashba
spin-orbit interaction has been investigated by means of a
nonequilibrium Green's functions approach. We have studied the
intrinsic and disorder-mediated contributions to spin-Hall conductivity
considering both the short- and long-range
hole-disorder scatterings. For short-range collisions, the
disorder-mediated contribution vanishes. When impurity scattering becomes
long-ranged, however, the disorder-mediated spin-Hall conductivity
becomes finite and can change the sign. It is negative for low HH
density and positive for high density. The temperature dependence
of disorder-mediated and total spin-Hall conductivity has also
been analyzed. The total spin-Hall conductivity always descends
with rising temperature at different HH densities, whereas the
behaviors of the disorder-mediated one versus temperature
become hole-density-related. A comparison with recent experiment
indicates that the observed spin-Hall effect is probably a result
of the contributions from both the intrinsic and disorder-mediated
processes.

\begin{acknowledgments}
The authors would like to thank
S. C. Zhang, S. Murakami, M. W. Wu, W. Xu, W. S. Liu, and Y. Chen
for useful discussions. Also, we are grateful to B. I. Halperin
and his colleagues for pointing out typographical errors in the previous
versions of this paper. This work was supported by projects of the National
Science Foundation of China and the Shanghai Municipal Commission of Science
and Technology, and by the Youth Scientific Research Startup
Funds of SJTU.
\end{acknowledgments}

\end{document}